\pdfoutput=1
\documentclass[aps,prl,print,superscriptaddress,twocolumn]{revtex4-1}
\usepackage{amsmath}% needed for subequations
\usepackage{graphicx}
\usepackage{siunitx}
\usepackage[colorlinks=true, linkcolor=blue, citecolor=blue, urlcolor=blue]{hyperref}

\begin{document}

\title{Long Range Magnetic order stabilized by acceptors}

\author{Xiaodong Zhang}
\affiliation{Department of Physics, Chinese University of Hong Kong, Hong Kong SAR, China}
%\affiliation{Department of Physics, Northwest University, Xi'an, 710069, China}

%\author{Yiou Zhang}
%\affiliation{Department of Physics, Chinese University of Hong Kong, Hong Kong SAR, China}
%
%\author{Kinfai Tse}
%\affiliation{Department of Physics, Chinese University of Hong Kong, Hong Kong SAR, China}
%
%\author{Bei Deng}
%\affiliation{Department of Physics, Chinese University of Hong Kong, Hong Kong SAR, China}

\author{Jingzhao Zhang}
\affiliation{Department of Physics, Chinese University of Hong Kong, Hong Kong SAR, China}

\author{Shengbai Zhang}
\affiliation{Department of Physics, Applied Physics, and Astronomy, Rensselaer Polytechnic Institute, Troy, New York 12180, United States}

\author{Junyi Zhu}\email{jyzhu@phy.cuhk.edu.hk}
\affiliation{Department of Physics, Chinese University of Hong Kong, Hong Kong SAR, China}

\begin{abstract}
Tuning magnetic order in magnetic semiconductors is a long sought goal. A proper concentration of acceptors can dramatically suppress local magnetic order in favor of the long one. Using Mn and an acceptor codoped LiZnAs as an example, we demonstrate, by first-principles calculation, the emergence of a long-range magnetic order. This intriguing phenomenon can be understood from an interplay between an acceptor-free magnetism and a band coupling magnetism. Our observation thus lays the ground for a precise control of the magnetic order in future spintronic devices.
\end{abstract}
\maketitle
\par

%\section*{Introduction}
For decades, diluted magnetic semiconductors (DMS) have attracted considerably interests for their unique magnetic properties and flexible tunabilities used in the information devices \cite{Dietl2000,Wolf2001,Ohno2000,Chiba2003,Chiba2008,Zhu2008,Peng2009,Dietl2014,Dietl2015,Matsukura2015,Deng2016,Qi2016,Deng2017}. Among the examples, a giant and tunneling magnetoresistance device was realized in trilayer structures of Mn doped GaAs \cite{Chung2010}. Magnetic memory cells using the Mn-doped GaAs have also been constructed, in which bit-writing can be processed by applying a magnetic field or spin-polarized electric field \cite{Figielski2007,Pappert2007,Mark2011}. However, the formation of inhomogeneous magnetic domains, originated from a short-range magnetic order, is a critical bottleneck to further developing the DMS technology \cite{Dunsiger2010,Nemec2013}. 

These domains are formed by an aggregation of the magnetic dopants \cite{Dietl2015,Mahadevan2005,Kuroda2007}, as a result of a short-range attractive interaction among them  \cite{Dietl2015}. It happens that a carrier doping can have a considerable effect on the aggregation. For example, Kuroda et al. showed that at certain concentrations of electron donors the aggregation of Cr in (Zn$_{1-x}$Cr$_x$)Te is enhanced. This phenomenon can be explained by an energy gain due to an electron-enhanced short-range magnetic interaction among the Cr atoms \cite{Silva2008}. In contrast, a long-range magnetic interaction can provide a thermodynamic driving force to reduce the undesirable phase separation \cite{Chang2013a,Chan2017}. Therefore, to overcome the bottleneck, an approach that can largely suppress the short-range magnetic order, while {\it simultaneously} enhancing the long-range one, is highly desirable, but unfortunately lacking.

The reason is because such an approach would be against the common belief that the long-range order is always weaker than the short-range one, because of a rapid decrease of the magnetic coupling strength with the distance between magnetic dopants. Such a belief has led to an impasse in advancing the study of DMS, which, in our view, is not flawless. For example, it is known that an acceptor-mediated magnetic order can work against the short-range magnetic order that dominates at the acceptor-free condition \cite{Dalpian2006}. In other words, intentionally-introduced acceptors could reduce the short-range magnetic coupling strength, whereby mitigating the undesirable magnetic phase separation. At the same time, the long-range magnetic order may stay or even be enhanced, as its response to the presence of the acceptors can be different from that of the short-range order. This reasoning raises the hope that at a proper magnetic dopant concentration and hole concentration, the system may suppress the phases that are in favor of short-range order by significantly increasing their formation energies.

In this Letter, we present a theory that reveals the interplay between different magnetic orders in the presence of hole doping. As an example, we consider magnetic orders in a Mn-doped LiZnAs, where Li$_{Zn}$ and/or V$_{Li}$ serve as the acceptors, to contrast with the ones in the absence of the acceptors. Using density functional theory calculations, we discover a long-range AFM order, which may be attributed to the stepping stone mechanism mediated by the magnetized Zn $d$ and As $p$ states. We also find, in line with the above discussions, that at a proper acceptor concentration, the short-range AFM order can be removed, while the long-range AFM order gives way to an FM order. The net effect is the stabilization of a long-range FM configuration. Our findings thus point to a new direction in rationally designing the DMS.

All calculations are performed using projected augmented-wave method \cite{Blochl1994} and density functional theory within generalized gradient approximation of Perdew-Burke-Ernzerhof \cite{Perdew2008} as implemented in VASP code \cite{Kresse1996}. The Mn doped LiZnAs is simulated by $2\times2\times2$ supercell (96 atoms). All atoms are relaxed with force tolerance of \SI{0.01}{eV/\AA}. A plane-wave cut-off energy of \SI{500}{eV} was used in all calculations. For Brillouin-zone integration, Monkhorst-Pack k-points grid of $4\times4\times4$ was employed. In order to consider strong correlation effect of transition metals, the LDA+U method \cite{Dudarev1998} is used. The Hubbard parameter $U$=\SI{3.5}{eV} and Hund rule exchange parameter $J$=\SI{0.6}{eV} are taken as suggested by Ref. \cite{Masek2007}. HSE06 calculations are also performed to check the accuracy of LDA+U method and found that results of LDA+U method are qualitatively consistent with results of HSE calculations. Convergence tests in respect to cell size, energy cutoffs, k-points, U, and force cutoffs have been performed.

\begin{figure}
\includegraphics*[width=7cm]{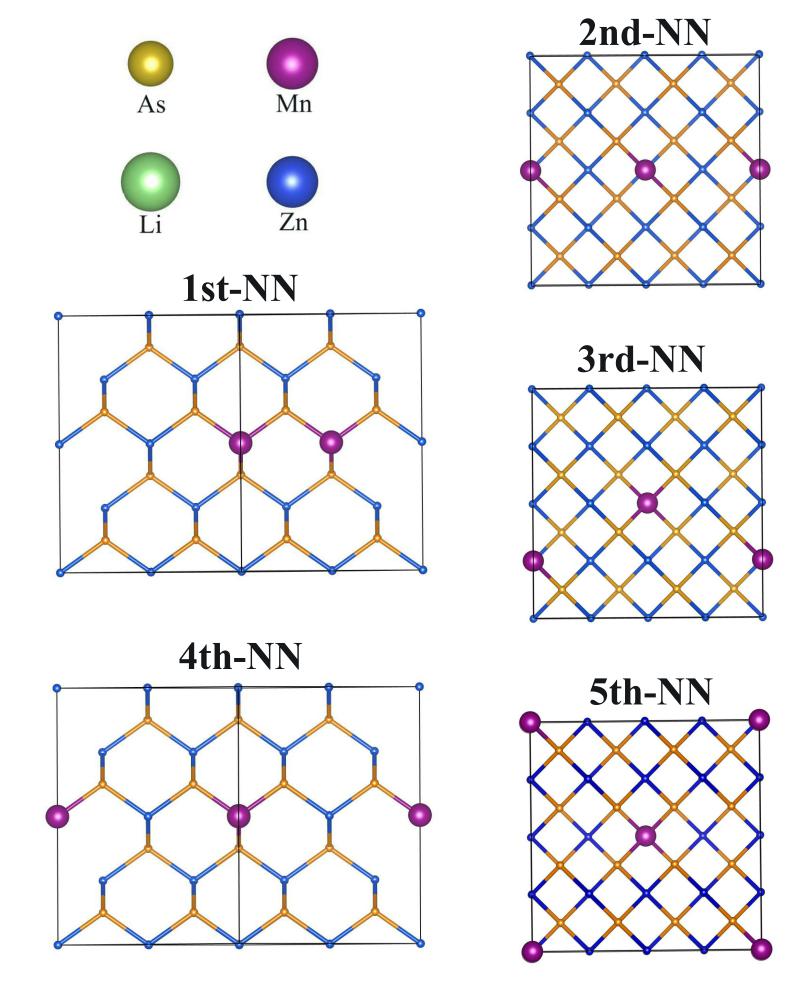}
\caption{Simulation cells with the dopant separation from the first nearest neighboring configuration to the 5th nearest neighboring one. The Li atoms are deleted for clarity. \label{fig:1}}
\end{figure}

  To simulate the short range and long range magnetism, we use $2\times2\times2$ supercell with two Mn atoms substituting two Zn atoms, which corresponding to Mn concentration of 6.25\%. Considering the symmetry of LiZnAs, there are 5 types of in-equivalent configurations, corresponding to first nearest neighbor (1st-NN) configuration to fifth nearest neighbor (5th-NN) configuration shown in Fig.\ref{fig:1}. 

  The formation energy of Mn doped LiZnAs is defined as
\begin{equation}
E_f = E(doped) - E(undoped)+ n_{Zn}\mu_{Zn} - n_{Mn}\mu_{Mn}
\end{equation}
where  $\mu_{Zn}$ and $\mu_{Mn}$  are chemical potential of Zn and Mn, respectively. And $n_{Zn}$ and $n_{Mn}$ are number of Zn and Mn, respectively. In the comparison among different magnetic configurations, these two chemical potentials will be cancelled.

\begin{figure}
\includegraphics*[width=7cm]{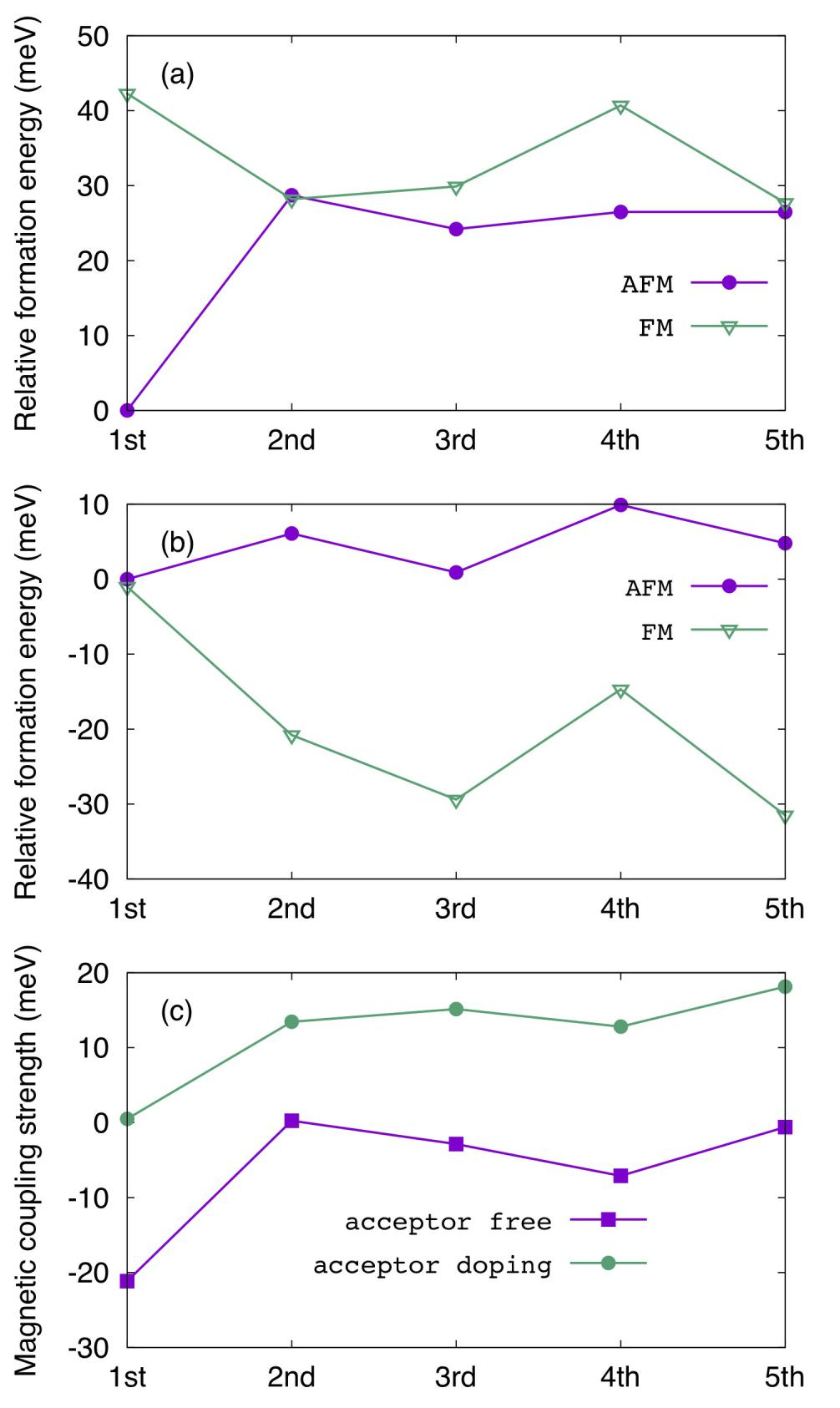}
\caption{(a) Relative formation energy of Mn dopants at different nearest neighboring sites  for acceptor free case; (b) that energy for the acceptor doping case; (c) magnetic coupling strength of different configurations.\label{fig:2}}
\end{figure}

%\section{Results and discussions}  
Firstly, we calculated the relative formation energy without acceptors as a function of different neighboring configurations, as shown in Fig.\ref{fig:2}(a). Here, the relative formation energy is calculated in reference to an AFM state of the first nearest neighboring (1st-NN) sites, which is the most stable configuration. 

Next, we introduced acceptors by replacing Zn atoms with Li atoms. Various Zn sites have been checked and we found that the most stable configuration is obtained by replacing Zn sites that are the nearest neighbors to the magnetic dopants. Still, the total energies are similar for different replacement sites. Also, we calculated the formation energy of the Mn pairs as a function of different doping configuration. To our surprise, the most stable configuration is the FM state with magnetic dopant atoms occupying the fifth nearest neighboring (5th-NN) sites and the 1st-NN sites becomes the most unstable Fig.\ref{fig:2}(b). This discovery strongly suggests that with the introduction of acceptors, the local magnetic dopants clustering is largely hindered and the long range order of magnetic dopants emerges.

To further understand this dramatic change in the relative stabilities of different configurations, we calculated the magnetic coupling strength of each configuration, as shown in Fig.\ref{fig:2}(c). The magnetic coupling strength is defined as a half of the difference between the AFM and FM state. The calculated results demonstrate that the first nearest neighboring sites prefer a AFM state for acceptor free case. However, when acceptor are introduced, the magnetic coupling strength decreases to almost zero. When the distance between the dopant pairs becomes longer, the magnetic order changed from AFM or non-magnetic to FM. This discovery is different from the ``common sense" belief that magnetic coupling strength is the strongest among the neighboring sites and decays very fast when the distance between dopant atoms increases.

In order to understand how acceptor doping changes the magnetic interaction and further influence the relative formation energy, we firstly need to understand the magnetic interaction in acceptor free case. 

In acceptor free case, the magnetic interaction of 1st-NN configuration can be explained by the superexchange theory, which suggests that for the half occupied $d$ state of Mn, the electron hopping from the As $p$ state strongly favors AFM coupling \cite{Anderson1950,Goodenough1955, Kanamori1959}. Still, the magnetic coupling between the third or fourth nearest neighbors are nonzero. A similar magnetic order has been discovered in Cr doped Bi$_2$Te$_3$ and Sb$_2$Te$_3$ system, where an antibonding state derived from the s lone pair on stepping stone Bi atoms plays a critical role for the long range magnetic order. However, in this system, there lacks such $s$ lone pair state.

\begin{figure}
\includegraphics*[width=6cm]{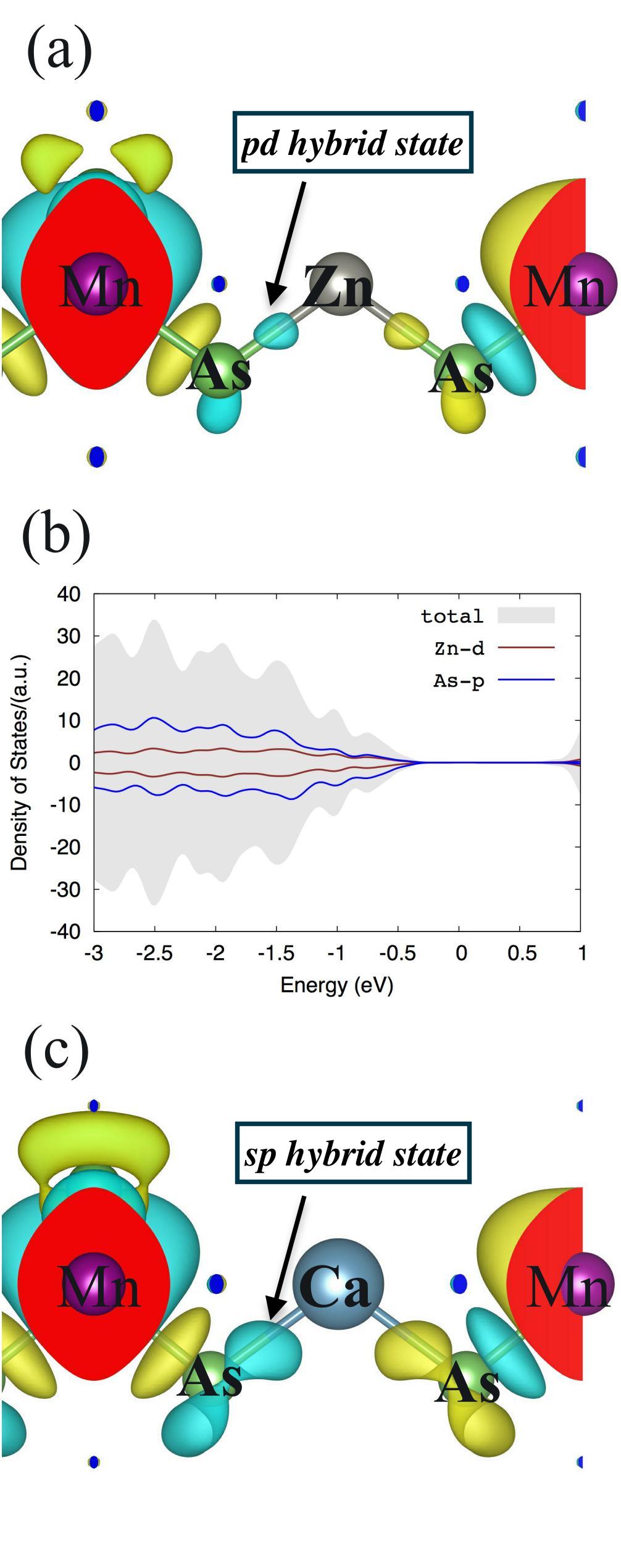}
\caption{ (a) Local spin density near dopant in Mn doped LiZnAs (b) Projected density of states (PDOS) (c) Local spin density near dopant in Mn,Ca codoped LiZnAs. \label{fig:3}}
\end{figure}

 To understand the long range magnetic coupling mechanism, we calculated the spin texture, as shown in Fig.\ref{fig:3}(a). We found that spin density exists near the center of As and Zn bond, demonstrating a covalent nature. The coupling between the As-$p$ state and Zn-$d$ state is clearly demonstrated in projected density of states, as shown in Fig.\ref{fig:3}(b). The electron hopping among the spin polarized covalent states near the center of the two As and Zn bonds on the chain lowers the total energy and enhances the long range correlation between the magnetic dopants.
 
\begin{figure}
\includegraphics*[width=7cm]{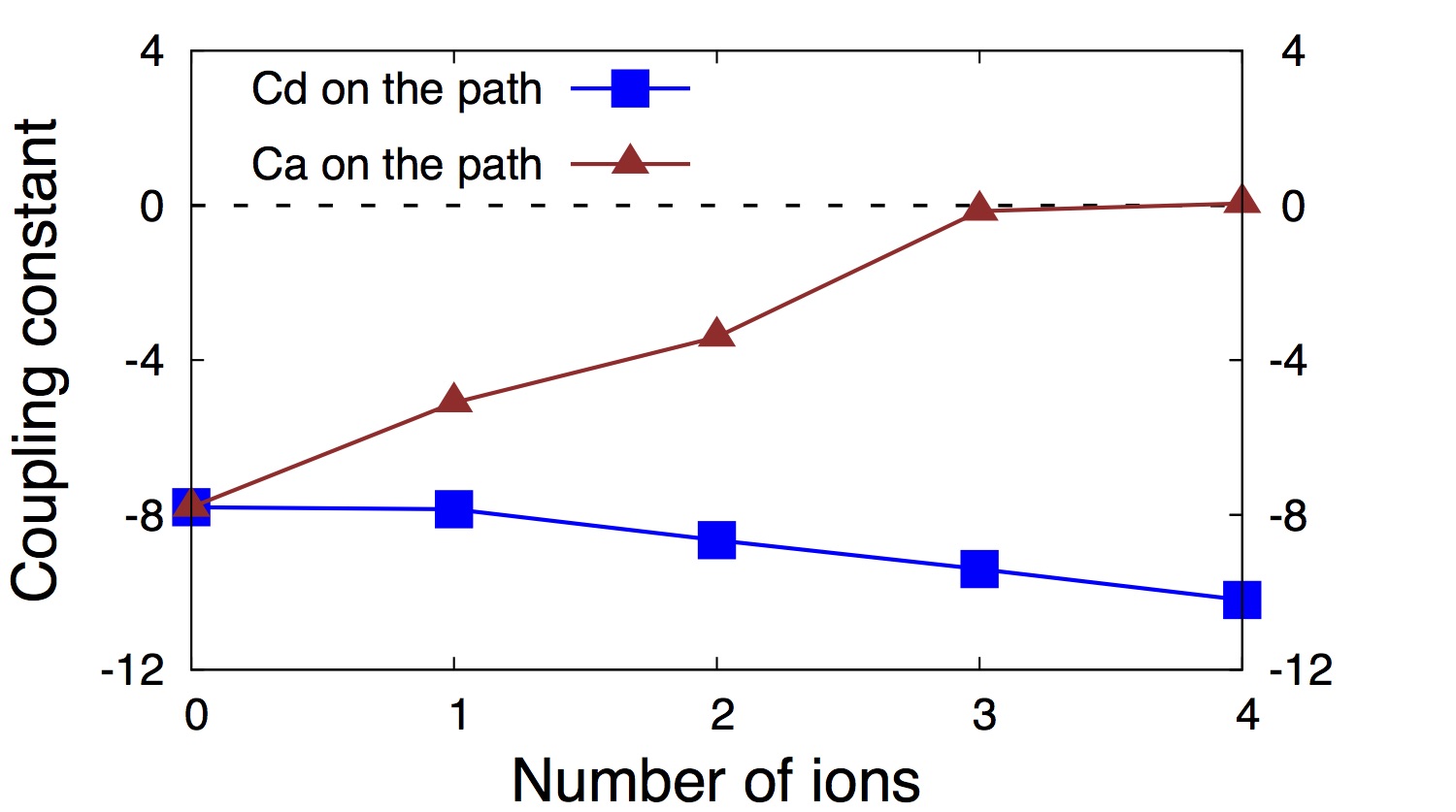}
\caption{Changes of the coupling strength with respect to the replaced atoms.\label{fig:4}}
\end{figure}

Further, we substitute Zn site in the Mn-As-Zn-As-Mn chain by Ca or Cd atoms in order to investigate the role of Zn for such long range magnetic interaction. We found that the spin density becomes localized around As atoms. This is due to the high ionicity of the Ca and As bond, as shown in Fig.\ref{fig:3}(c). Note that, there is no $d$ orbital in Ca$^{2+}$ and the $s$ electron are mostly transferred to the orbitals near the As atom. The electron hopping between the orbitals near different As atoms on the chain is largely hindered because the overlap of orbitals almost disappeared (refer to supplementary materials for details). As expected, we found that the magnetic coupling strength is almost zero when Zn sites in the four Mn-As-Zn-As-Mn chains surrounding one Mn atom are substituted by four Ca atoms (Fig.\ref{fig:4}). Further, we also substitute four Zn atoms by Cd atoms. We found that the magnetic coupling preserves because the coupling between the d orbitals of Cd and the p orbitals of As is covalent and near the center of the As and Cd bond. Therefore, the electron hopping among these states can lower the total energy and enhance the long range correlation between Mn dopants. The large coupling strength in Cd doping case is due to the enhanced $p$-$d$ coupling strength between Cd and As, since the d orbital of Cd is higher than that of Zn.
 
The above analysis demonstrates that the covalent nature of $p$-$d$ hybridized orbital is the direct reason for such long range magnetic interactions. However, such enhancement in the magnetic coupling is not limited to the $p$-$d$ coupling mechanism.

Next, we'll investigate the interplay between acceptor mediated magnetism and the intrinsic long range magnetism. We first studied acceptor doping. Band coupling model \cite{Dalpian2006} suggests that the exchange coupling strength is sensitive to the position of $d$ level, relative to the VBM. Under the crystal field of $T_d$ symmetry, the Mn $d$ orbitals are split into lower $e_g$ states and higher $t_{2g}$ states. All the five $d$ electrons of Mn will occupy these states, which are the majority spin states below the VBM. The minority spin states are all empty and above the VBM. The original band coupling model \cite{Dalpian2006} proposed that the energy difference between FM and AFM phase is
\begin{equation}
\Delta E_{FM-AFM} = -\alpha n_{h} (\Delta_{pd}^1 + \Delta_{pd}^2) + 6 \Delta_{dd}^{1,2}
\label{eq:origin}
\end{equation}
where $\alpha$ is a parameter related to the localization of hole (acceptor) states and Mn-Mn distance; $n_h$ is the hole (acceptor) density; the $\Delta_{pd}^1$ and $\Delta_{pd}^2$ are contributions from acceptors and described by Zener's model \cite{Zener1951,Dietl1997,Dietl2000}; the $\Delta_{dd}^{1,2}$ is coming from intrinsic magnetic interaction. More details about band coupling model and parameters can be found in Ref.\cite{Dalpian2006}. However, this energy difference is not a function of the distance between the dopants. To include the distance as an important variable in the long range magnetic investigation, we slightly modified the above equation, as listed below,

\begin{eqnarray}
\Delta E_{FM-AFM} &= -\alpha(R) n_{h} (\Delta_{pd}^1 + \Delta_{pd}^2) + 6\beta(R) \Delta_{dd}^{1,2}(n_h) \nonumber \\
&=n_h J_{acceptors}(R) + J_{intrinsic}(R,n_h)
\label{eq:revised}
\end{eqnarray} 

where R is the distance between the two magnetic atoms; $n_h$,  $\Delta_{pd}^1$,  $\Delta_{pd}^2$, and  $\Delta_{dd}^{1,2}$ are the same as Eq.\ref{eq:origin}. $\alpha(R)$ and $\beta(R)$ are magnetic interaction parameters as functions of the distances, based on the acceptor mediated mechanism, and the intrinsic long range mechanism, respectively. Usually, the decay of $\beta(R)$ is much faster than $\alpha(R)$ because acceptor state is very delocalized and the interaction range can be very long \cite{PeterYu2010}. The decay of $\beta(R)$ is sensitive to symmetry of bonding, because the intrinsic coupling is mediated by the electron hopping among the magnetized orbitals of the host material. acceptor mediated magnetic interaction $J_{acceptor} = -\alpha(R)(\Delta_{pd}^1 + \Delta_{pd}^2)$ is usually negative while intrinsic magnetic interaction $J_{intrinsic} = 6\beta\Delta_{dd}^{1,2}(n_h)$ is usually positive. The  $J_{intrinsic}(R,n_h)$ is not only dependent on distance between the two magnetic atoms, but also influenced by the acceptor density. We further checked the $J_{intrinsic}$ at different dopant sites and found that the interaction still exists for first nearest neighbors upon acceptor doping. However, the long range intrinsic interaction is destroyed by acceptors. This is probably due to the change of electron occupation in the stepping stone state, which is more sensitive to the acceptor doping than the As-$p$ state that mediates the super exchange mechanism. This difference directly leads to the significant different magnetic order upon acceptor incorporation at different magnetic doping sites. A more complete picture can be achieved by strict analysis of many body effects in the future, which is out of the scope of this paper [refer to Supplementary materials for details].
  
 Since the first term in Eq.\ref{eq:revised} is dependent upon the acceptor density, it is possible to tune the magnetic coupling by changing the acceptor concentration. If the sign of  $J_{acceptor}$ and  that of $J_{intrinsic}$ are opposite, a proper acceptor density will result in zero magnetic interaction for short range configurations (1st-NN). 
 
 In our simulation cell, when we introduce one acceptor by removing one electron from the system, the magnetic coupling becomes almost zero on the first nearest neighboring site. When we introduce a Li substitutional defect on a Zn atom that is close to the Mn atom, or when we remove one Li atom that is close to the Mn atom, the magnetism on the nearest neighboring site configuration also disappeared. Therefore, these three calculations confirmed that the acceptor cancels the magnetism on nearest neighboring configurations.

Further, we checked the magnetic order on other neighboring sites. For acceptor free case, the 2nd-NN and the 5th-NN configuration yield almost zero magnetic coupling. Despite the large difference of Mn-Mn distance between 2nd-NN (\SI{5.94}{\AA}) and 5th-NN (\SI{10.28}{\AA}), the magnetic coupling strength of 2nd-NN (\SI{13.5}{meV}) and that of 5th-NN (\SI{15.1}{meV}) are similar. These results suggest that acceptor induced magnetic interaction is almost a constant shift in different neighboring sites, different from the fast decay nature under acceptor free condition, as shown in Fig.\ref{fig:2}(c). This difference largely cancels the local magnetic interaction and a long range magnetic interaction emerges. 

Based on all these calculations and analysis, a  strategy on tuning the magnetism of different sites can be proposed. If the short range magnetic interaction under acceptor free condition is different from the acceptor mediated magnetic interaction, it's possible to incorporate a proper amount of acceptors to largely destroy the short range magnetism. As a result, the short range magnetic configuration become unstable and long range magnetism emerges. Such long range magnetic interaction stabilizes long range configurations and results in a long range magnetic ordering phase, which can be the global minimum. As shown in Fig.\ref{fig:2}(b), the formation energies of the configurations with three or more atoms that separate the dopant atoms (from 2nd-NN to 5th-NN) are all lower than that of the 1st-NN configuration when a acceptor is doped.

This is the very first time, a sensitive relationship between the stability of short range \emph{vs.} long range magnetic order and the concentration of the acceptors is discovered. With a proper amount of co-dopants are incorporated, the long range magnetic order can suppress the short range one and become stable. \textbf{Such stability is very important during the growth of DMS, because once the spinodal decomposition is formed due to the strong short range magnetic coupling, it's often kinetically forbidden to change the magnetic coupling into long range ones via post annealing techniques.} These results are also consistent with early experimental discoveries, which suggest that the formation of nanocrystal (results of spinodal decomposition) can be tuned by acceptors or donors in Cr doped ZnTe \cite{Kuroda2007}. Therefore, we expect our strategy should be general in various transition metals doped in DMS and may lead to  discoveries of class of magnetic materials. Our discovery also strongly suggests that it's usually naive to use short range magnetic order to represent the long range one. To achieve a complete picture of magnetic order upon different dopant to dopant separations, various doping configurations have to be tested to guarantee the correct results.

%\section{Conclusions} 
In summary, we observe, based on density functional theory calculations, that the long-range AFM order in an acceptor-free Mn-doped LiZnAs is mediated by magnetized $pd$ hybridized states. While here the short-range magnetic order can be largely suppressed by a proper acceptor codoping with Mn, it happens that the long-range AFM order also simultaneously ceases, giving way to a long-range FM order. Such a long-range order is the long sought goal in the DMS study. Our codoping strategy should be general, whose applications to other DMS may lead to a discovery of materials with stable long-range magnetism. It may also serve as a means to control magnetic order between the FM and AFM phases. 
 \section*{Acknowledgements}
XZ, JZZ, and JYZ are grateful for the financial support of Chinese University of Hong Kong (CUHK) (Grant No.4053084), University Grants Committee of Hong Kong (Grant No.
24300814), and start-up funding of CUHK. SBZ acknowledges the support by the U.S. DOE Grant No. DESC0002623.

%\bibliographystyle{apsrev4-1}
%\bibliography{ref}

%
\end{document}